\documentclass[aps,superscriptaddress,a4paper]{revtex4}
\usepackage[colorlinks=true, pdfstartview=FitV, linkcolor=blue, citecolor=red, urlcolor=magenta]{hyperref}
\usepackage{graphicx}
\usepackage{latexsym}
\usepackage{amsmath}
\usepackage{amsfonts}
\usepackage{amssymb}
\usepackage{verbatim}

\newcommand{\be}{\begin{equation}}
\newcommand{\ee}{\end{equation}}
\newcommand{\bea}{\begin{eqnarray}}
\newcommand{\eea}{\end{eqnarray}}

\newcommand{\ben}{\begin{eqnarray}}
\newcommand{\een}{\end{eqnarray}}

\newcommand{\bes}{\begin{subequations}}
\newcommand{\ees}{\end{subequations}}

\begin{document}

\title{Quantum-corrected two-dimensional Horava-Lifshitz black hole entropy}

\author{M. A. Anacleto}
\email{anacleto@df.ufcg.edu.br}
\affiliation{Departamento de F\'\i sica,
Universidade Federal de Campina Grande, Caixa Postal 10071,
58429-900  Campina Grande, Para\'\i ba,
Brazil
}

\author{D. Bazeia}
\email{bazeia@fisica.ufpb.br}
\affiliation{Departamento de F\'\i sica, Universidade Federal da Para\'\i ba, 58051-970 Jo\~ao Pessoa, PB, Brazil}

\author{F. A. Brito}
\email{fabrito@df.ufcg.edu.br}
\affiliation{Departamento de F\'\i sica,
Universidade Federal de Campina Grande, Caixa Postal 10071,
58429-900  Campina Grande, Para\'\i ba,
Brazil
}\affiliation{Departamento de F\'\i sica, Universidade Federal da Para\'\i ba, 58051-970 Jo\~ao Pessoa, PB, Brazil}

\author{J. C. Mota-Silva}
\email{julio{\_}cmota@yahoo.com.br}
\affiliation{Departamento de F\'\i sica, Universidade Federal da Para\'\i ba, 58051-970 Jo\~ao Pessoa, PB, Brazil}

\begin{abstract}
In this paper we focus on the Halmiton-Jacobi method to determine  {several thermodynamic quantities such as the temperature, entropy and specific heat} of two-dimensional Horava-Lifshitz black holes by using the generalized uncertainty principles (GUP). We also address the product of horizons, mainly concerning the event, Cauchy, cosmological and virtual horizons.
\end{abstract}
\pacs{XX.XX, YY.YY} \maketitle

%\vspace{1cm}

\section{Introduction}
Theories of gravity in two spacetime dimensions has received much attention in the 
literature~\cite{mann, twodimentional} 
and  provides an excellent theoretical laboratory for understanding issues relevant to quantum gravity.
Such theories in recent years has presented a very rich structure and an interesting relationship with
conformal field theory~\cite{Polyakov87}, the Liouville model~\cite{Teitelboim}, random lattice models~\cite{Polyakov88}, and sigma models~\cite{leblanc88}. 
Formally, it has similarity to four-dimensional general relativity. In fact, the solution of this theory has a nontrivial event horizon structure that enables the existence of black holes in two spacetime dimensions. 
Recently, a new theory of gravity has been presented by Horava in~\cite{Horava}, and this is now the well-known Horava-Lifshitz (HL) gravity. Many aspects of the theory have been considered in the literature \cite{horava2}. 
In~\cite{geraldo}  was proposed a new Horava-Lifshitz black hole solution in  two dimensions  in the slow varying dilatonic field regime.

The main objective of the present study is to address the issues of quantum-corrected entropy in the two-dimensional HL black hole. A semiclassical approach considering the Hawking radiation as a tunneling phenomenon across the horizon has been proposed in Refs.~\cite{Parikh, ec.vagenas-plb559}, in addition to the Hamilton-Jacobi method~\cite{SP} to determine the Hawking radiation and the entropy of black holes.  The tunneling formalism has also been applied to HL gravity, for instance, in~\cite{Eune:2010kx,Liu:2011gya}.
In~\cite{Parikh, Jiang} the method of radial null geodesic was used by the authors for calculating the Hawking temperature. In Ref.~\cite{Banerjee} applying the tunneling formalism has been investigated Hawking radiation considering  self-gravitation and back reaction effects. 
{It was presented in~\cite{Singleton:2010gz,Singleton:2013ama} the information loss paradox in the WKB/tunneling picture of Hawking radiation considering the back reaction effects.}
More recently, using this formalism the back reaction effects for self-dual black hole has also been investigated~\cite{Silva12}. 
It was calculated in~\cite{Majumder:afa} the quantum-corrected Hawking temperature and entropy of a Schwarzschild black hole  considering the effects of the generalized uncertainty principle (GUP) in the tunneling formalism.
Moreover,  using the Hamilton-Jacobi version of the tunneling formalism were investigated the Hawking radiation for acoustic black hole~\cite{Becar:2010zza} and in~\cite{Anacleto:2015mma} has been discussed the thermodynamic properties of self-dual black holes and 
non-commutative BTZ black hole~\cite{Anacleto:2015kca, Ovgun:2015box}.
It was analyzed in~\cite{Faizal:2014tea} the corrections for the thermodynamics of black holes
assuming that the GUP corrected entropy-area relation is universal for all black objects.

A lot of work has been proposed in the literature in order to understand the quantum aspects of the black hole entropy  --- see for instance ~\cite{Wilczek, Magan:2014dwa,Solodukhin:2011gn,mhorizon}. 
In Ref.~\cite{Kaul} the authors have shown that the quantum corrections to the Bekenstein-Hawking entropy 
is logarithmic and  dependent of the area.
Furthermore, it was obtained in Ref.~\cite{Carlip:2000nv}  an additional correction term to the entropy that depends on conserved charges.
In addition, in Ref.~\cite{Rinaldi} using the brick-wall method was investigated entropy of acoustic black hole in two dimensions. 
However, when determining the entropy by this method an ultraviolet cut-off must be inserted in the calculations to eliminate the divergence of states density near black hole horizon. 
On the other hand, considering models in which the Heisenberg uncertainty relation is modified the divergence that arise in the brick-wall model are eliminated~\cite{Brustein, XLi, KN, ABPS, Zhao, Zhang, review}.
For example, in one-dimensional space, we have the following modified Heisenberg uncertainty relation
\[
\Delta x \Delta p \geq \frac{\hbar}{2}\left[1 + \alpha^{2}(\Delta p)^{2}\right],
%\label{int1}
\]
where $ \Delta x $ and $\Delta p $ are uncertainties for position and
momentum, respectively, and $ \alpha $ is a positive constant which is independent of $ \Delta x $ and $\Delta p $.
{The thermodynamic properties of black holes are modified due to the GUP~\citep{Adler:2001vs}. 
See for instance~\cite{Nicolini:2008aj}
which reviews how to approach spacetime noncommutativity that leads to an effective GUP to quantum gravity.}

In order to obtain a change in the temperature of the black hole first the bound on the maximum momentum in GUP is identified with a bound on maximum energy of the system and the uncertainty in the position can be taken to be proportional to the radius of the event horizon of the black hole~\cite{Gangopadhyay:2015zma}.

In this paper, inspired by all of these previous works, we shall mainly focus on new Horava-Lifshitz black hole solutions in  two dimensions and  the Hamilton-Jacobi method in order to determine the temperature and the entropy of a black hole using the GUP.

\section{Tunneling formalism for 2d HL black hole}
\label{tunel}

In this section, we will use the tunneling formalism to derive the Hawking temperature for a two-dimensional HL black hole. In our calculations we assume that the classical action satisfies the relativistic Hamilton-Jacobi equation to leading order in the energy. The metric in (Arnowitt-Deser-Misner) ADM decomposition is
\begin{eqnarray}
ds^2=-N^2dt^2+g_{ij}(dx^i+N^idt)(dx^j+N^jdt)
\end{eqnarray}
with an anisotropic scaling between space and time, $t\to b^{-z}t,\, x^i\to b^{-1}x^i,\, i=1,2,...,D$. The power-counting renormalizability requires $z\geq D$. In 1+1 dimensions this means $z\geq 1$ {that we shall assume $z=1$, i.e, 
infrared regime. Despite of this choice, the HL gravity does not coincide with Einstein gravity in general --- see~\cite{Horava,Liu:2011gya} for further discussions on IR issues in HL gravity. }
Now, by using the gauge $N_1=0$ and considering $g_{ij}\equiv g_{11}=N^{-2}$ we have
\begin{eqnarray}
ds^{2} = -f(x)dt^{2} + f(x)^{-1}dx^{2},
\label{1.21}
\end{eqnarray}
where we have redefined $N^2\equiv f(x)$. The two-dimensional HL black hole solutions are explicitly given by~\cite{geraldo}
\begin{eqnarray}
f(x) = 2C_{2} + \frac{A}{\eta}x^{2} - 2C_{1}x + \frac{B}{\eta x} + \frac{C}{3 \eta x^{2}},
\label{1.1}
\end{eqnarray}
where $\eta$ is related to the non-projectable version of HL gravity. In the sequel we consider the Klein-Gordon (KG) equation
\begin{eqnarray}
\hbar^{2}g^{\mu\nu}\nabla_{\mu}\nabla_{\nu}\phi - m^{2}\phi=0.
\label{1.22}
\end{eqnarray}
Notice that the KG equation preserves its usual form because we have fixed $z=1$ and the gauge $N_1=0$ as discussed above. Similar considerations in higher dimensions have been considered both in arbitrary \cite{Eune:2010kx} and fixed $z=1$ (IR regime) \cite{Liu:2011gya}.
Now considering the metric (\ref{1.21}), we have
\begin{eqnarray}
-\partial^{2}_{t}\phi  + f(x)^{2}\partial^{2}_{x}\phi + \frac{1}{2}f(x)^{2\prime}\partial_{x}\phi - \frac{m^{2}}{\hbar}f(x)\phi = 0.
\label{1.23}
\end{eqnarray}
Next we apply the WKB approximation to $ \phi $ given by
\begin{eqnarray}
\phi(x,t) = \exp\left[-\frac{i}{\hbar}{\cal I}(x,t)\right],
\label{1.24}
\end{eqnarray}  
and to the lowest order in $ \hbar $, we have
%Then, using tha real parte of Eq. (\ref{1.23}), we have
\begin{eqnarray}
(\partial_{t}{\cal I})^{2} - f(x)^{2}(\partial_{x}{\cal I})^{2} - m^{2}f(x) = 0.
\label{1.25}
\end{eqnarray}  
Because of the symmetries of the metric, we can write a solution to $ {\cal I}(x,t) $ in the form
\begin{eqnarray}
{\cal I}(x,t) = -\omega t + W(r),
\label{1.26}
\end{eqnarray}
where for $W(r)$ we have
\begin{eqnarray}
W = \int{\frac{dx}{f(x)}\sqrt{\omega^{2} - m^{2}f(x)}}.
\label{1.27}
\end{eqnarray}
At this point, we can apply near the horizon the following approximation
%In this point, by taking the near horizon approximation  
\begin{eqnarray}
f(x) = f(x^{+}_{h}) + f^{\prime}(x^{+}_{h})(x - x^{+}_{h})+\cdots 
\label{1.28}
\end{eqnarray}
In this way, for the spatial part of the action function we find
\begin{eqnarray}
W &=& \int{\frac{dx}{f^{\prime}(x^{+}_{h})}
\frac{\sqrt{\omega^{2} - m^{2}f^{\prime}(x^{+}_{h})(x - x^{+}_{h})}}{(x - x^{+}_{h})}}
%\nonumber\\
=\frac{2\pi i \omega}{f^{\prime}(x^{+}_{h})}.
\label{1.29}
\end{eqnarray}
Therefore, the tunneling probability for a particle with energy $\omega$ is given by
\begin{eqnarray}
\Gamma \cong \exp[-2 Im {\cal I}]=\exp\left[-\frac{4\pi\omega}{f^{\prime}(x^{+}_{h})}\right].
\label{1.30}
\end{eqnarray}
Thus, comparing Eq. (\ref{1.30}) with the Boltzmann factor $(e^{-{\omega}/{T}})$, we obtain the  general  Hawking temperature formula for the black hole solution (\ref{1.1})
\begin{eqnarray}
T_{HL} = \frac{\omega}{2 Im {\cal I}}=\frac{f^{\prime}(x^{+}_{h})}{4\pi}.
\label{1.31}
\end{eqnarray}  
Below we shall mainly consider three cases. 

\subsection{First Case: Schwarzschild-like black hole}

In this case we consider $C_{1} \neq 0$, $C_{2} \neq 0$, $B \neq 0$ and $A = C = 0$ into Eq. (\ref{1.1})
and the metric becomes 
\begin{eqnarray}
f(x) = 2C_{2} - 2C_{1}x + \frac{B}{\eta x}.
\label{1.2}
\end{eqnarray}
The event horizons can be obtained at $ f(x)=0 $, such that we have
\begin{eqnarray}
x^{\pm}_{h} = \frac{C_{2}}{2C_{1}} \pm \sqrt{\frac{C^{2}_{2}}{4C^{2}_{1}} + \frac{B}{2C_{1}\eta}}.
\label{1.3}
\end{eqnarray}
{For special case $C_{2} = 0$, $C_{1} = -M$ and $B = -4M\Lambda^2$  (where $ \Lambda $ is a parameter with dimension of length), the horizons are }
\begin{eqnarray}
x^{\pm}_{h} = \pm\Lambda\sqrt{\frac{2}{\eta}}.
\label{1.5}
\end{eqnarray}
Thus, considering Eq. (\ref{1.2}) and substituting Eq. (\ref{1.5}) into Eq. (\ref{1.31}) we obtain the temperature given by 
\begin{eqnarray}
T_{HL1} = \frac{M}{\pi}.
\label{1.6}
\end{eqnarray}
Since the radius of the horizon in equation (\ref{1.5}) is independent of the mass $M$,
 the Hawking temperature is directly proportional to the mass parameter $M$, contrary to the case in four dimensions where Hawking temperature is inversely proportional to the mass parameter $M$.
To compute the entropy, we use 
\begin{eqnarray}
S_{HL1}  = \int{\frac{dM}{T_{HL1}}},
\label{1.71}
\end{eqnarray}
that substituting Eq.~(\ref{1.6}) into Eq.~(\ref{1.71}), we find
\begin{eqnarray}
S_{HL1}  = \pi\ln\left(\frac{M}{M_{0}}\right)=\pi\ln\left(\pi M^2\right)-\pi\ln\left(\pi{MM_{0}}\right).
\label{1.7}
\end{eqnarray}
This is the expected entropy in $(1+1)-$dimensional black holes \cite{mann}. 
{Furthermore, the first term in equation (\ref{1.7}) resembles a correction term which is of type $ \ln(A/4)=\ln(4\pi M^2) $ for the entropy of black holes in four dimensions.}
{Notice that the meaning of entropy for two-dimensional black holes is different of higher dimensional cases. This is because the event horizon is a point, i.e. has no area. However,  it still {enjoys} the thermodynamic relationship \cite{Bekenstein}
\begin{eqnarray}
dM = TdS - \Phi dQ,
\label{bek}
\end{eqnarray} 
where $\Phi$ is the electric potential, and there is no angular momentum term. Thus, the horizon has its own associated temperature and entropy and we can use (\ref{bek}) to define the entropy \cite{mann}. Since the constant $M_{0}$ plays the role of a fundamental length, the thermodynamic properties of a two-dimensional black hole require this length \cite{mann}. This seems to  have the  characteristic of theories that breakdown at the semi-classical regime. So, a minimum measurable length implies a major revision of quantum physics \cite{review}. These approaches is precisely  the GUP, as we see in section \ref{sectiongup}.
}\\
{\indent Several other interesting thermodynamic quantities can also be found as follows. The specific heat is given by 
\begin{eqnarray}
C = T\frac{\partial S}{\partial T},
\label{definitionsh}
\end{eqnarray} 
that from Eqs. (\ref{1.6}) and (\ref{1.7}) reads
\begin{eqnarray}
C_{HL1} = \pi.
\label{sh1}
\end{eqnarray} 
\indent Furthermore, the Hawking temperature can be used to compute the emission rate. Let us assume that in the black hole the energy loss is dominated by photons \cite{tawfik}. Then,  using the Stefan-Boltzmann law in two-dimensional spacetime we have
\begin{eqnarray}
\frac{dM}{dt} \propto T^{2}.
\label{er1}
\end{eqnarray} 
Thus, the emission rate in this case is
\begin{eqnarray}
\frac{dM_{HL1}}{dt} \propto \frac{M^{2}}{\pi^{2}}.
\label{er2}
\end{eqnarray} 
}
\subsection{Second Case: Reissner-Nordstr\"om-like black hole}
{In this second case, we will make $B = C_{1} = C_{2} = 0$ and $C = -3Q^2\Lambda^2$ in Eq. (\ref{1.1}), so for the function $ f(x) $ we have }
\begin{eqnarray}
f(x) = \frac{A}{\eta}x^{2} - \frac{Q^2\Lambda^2}{\eta x^{2}},
\label{1.8}
\end{eqnarray}
{Now choosing $A = \Lambda^{-2}$ and $Q^2=M^{2}\Lambda^{2}$ (for an extreme-like case), 
we have that the event horizons are }
\begin{eqnarray}
x^{\pm}_{h} = \pm \left(\frac{Q^2\Lambda^2}{A}\right)^{\frac{1}{4}}
=\pm\Lambda\sqrt{M\Lambda} .
\label{1.9}
\end{eqnarray}
Using Eq. (\ref{1.31}) the temperature is given by
\begin{eqnarray}
T_{HL2} = \frac{1}{\pi\eta}\left({Q^2\Lambda^2 A^3}\right)^{\frac{1}{4}}
\label{1.10}
%\end{eqnarray}
%T_{HL2} 
=\frac{1}{\pi\eta}\sqrt{\frac{M}{\Lambda}}.
\end{eqnarray}
{Here, unlike the first case, the Hawking temperature is proportional to $ \sqrt{M} $.
For the entropy we find the following result}
\begin{eqnarray}
S_{HL2} =\int \frac{d M}{T_{HL2}}={2\pi\eta}\sqrt{M\Lambda}.
\label{1.11}
\end{eqnarray}
{Now following the same steps of the previous case we can also compute identical thermodynamic quantities. From Eqs. (\ref{1.10}) and (\ref{1.11}), the specific heat now reads
\begin{eqnarray}
C_{HL2} = 2\pi\eta\sqrt{M\Lambda},
\label{sh2}
\end{eqnarray}
and the emission rate is 
\begin{eqnarray}
\frac{dM_{HL2}}{dt} \propto \frac{1}{(\pi\eta)^{2}}\frac{M}{\Lambda}.
\label{er3}
\end{eqnarray}
} 
%{\bf Note that for this case, the entropy is proportional to $ \sqrt{M} $, which is a different result 
%to that obtained for the black hole entropy in two dimensions in the first case.}

\subsection{The AdS-Schwarzschild-like case}
In this case we consider $A \neq 0$, $B \neq 0$ and $C = C_{1} = C_{2} = 0$ into Eq. (\ref{1.1}) and the metric becomes 
\begin{eqnarray}
f(x) = \frac{A}{\eta}x^{2} + \frac{B}{\eta x}
\label{ads1}
\end{eqnarray}
{Now choosing $A =\Lambda^{-2}$ and $B=-4M\Lambda^2$, we get the real event horizon which is }
\begin{eqnarray}
x_{h} = \left(-\frac{B}{A}\right)^{\frac{1}{3}}
=\left(4M\Lambda^4\right)^{1/3}.
\label{ads2}
\end{eqnarray}
Using Eq. (\ref{1.31}) the temperature is given by
\begin{eqnarray}
T_{AdS-S} = \frac{3}{4\pi\eta}\left(-A^2 B\right)^{\frac{1}{3}}
=\frac{3}{4^{2/3}\pi\eta}\left(\frac{M}{\Lambda^2}\right)^{1/3}.
\label{ads3}
\end{eqnarray}
For the entropy we find the following result
\begin{eqnarray}
S_{AdS-S} = {2^{1/3}\pi\eta}\left({M}{\Lambda}\right)^{2/3}.
\label{ads5}
\end{eqnarray}
%which has resemblance with the first case.
{Finally in the present case from Eqs. (\ref{ads3}) and (\ref{ads5}), we find the specific heat
\begin{eqnarray}
C_{AdS-S} = 2^{\frac{4}{3}}\pi\eta\Lambda \frac{M}{(M\Lambda)^{\frac{1}{3}}} .
\label{sh3}
\end{eqnarray}
The emission rate now reads
\begin{eqnarray}
\frac{dM_{AdS-S}}{dt} \propto \frac{9}{4^{4/3}\pi^{2}\eta^{2}}\left(\frac{M}{\Lambda^2}\right)^{2/3}.
\label{er4}
\end{eqnarray}
Notice that, except in the first example, all the thermodynamic quantities go to zero as $M\to0$. This phenomenon prevents the existence of black hole remnants \cite{Adler:2001vs,tawfik}. In the next section, we will find black holes with richer thermodynamic scenarios due to the GUP.}

\section{Quantum corrections to the entropy}
\label{sectiongup}
In this section, we will consider the GUP and we will apply the Hamilton-Jacobi method  in tunneling formalism 
to calculate the quantum-corrected Hawking temperature and entropy for a two-dimensional Horava-Lifshitz black hole. Hence, for the GUP we have~\cite{Brustein,XLi,KN,ABPS,Zhao,Zhang,review,Gangopadhyay:2015zma}
\begin{eqnarray}
\Delta x \Delta p \geq \hbar\left(1 - \alpha\frac{l_{p}}{\hbar}\Delta p + \frac{\alpha^{2}l^{2}_{p}}{\hbar^{2}}(\Delta p)^{2}\right),
\label{gup1}
\end{eqnarray}
where $\alpha$ is a dimensionless positive parameter and $l_{p}$ is the Planck length. %$l_{p} = \sqrt{\frac{\hbar G}{c^{3}}} = \frac{M_{p}G}{c^{2}} \approx 10^{-35}m$ in the Planck  lenght, $M_{p}$ is the Planck mass and $c$ is the velocity of light. 
We can still write the equation (\ref{gup1}) as follows
\begin{eqnarray}
\Delta p \geq \frac{(\Delta x + \alpha l_{p})}{2\alpha^{2}l^{2}_{p}}\left(1 - \sqrt{1 - \frac{4\alpha^{2}l^{2}_{p}}{\left(\Delta x + \alpha l_{p}\right)^{2}}}\right),
\label{gup2}
\end{eqnarray}
where we have chosen the negative sign and $ \hbar=1 $. 
Here since $\frac{l_{p}}{\Delta x}\ll 1$, the above equation can be expanded into Taylor series
\begin{eqnarray}
\Delta p \geq \frac{1}{\Delta x}\left[1 - \frac{\alpha {l_p} }{\Delta x} + \frac{2\alpha^{2}{l_p^2}}{(\Delta x)^{2}}+\cdots\right].
\label{gup3}
\end{eqnarray}
%By considering natural units $(c = k_{B} = \hbar = 1)$ and $l_{p} =1$, the generalized uncertainty principle becomes
%\begin{eqnarray}
%\Delta p\Delta x \geq 1.
%\label{gup3.1}
%\end{eqnarray}
{Now, with $ \hbar=1 $ the uncertainty
principle becomes $ \Delta x \Delta p \geq 1 $  and applying the saturated form of the uncertainty principle we have, $ \omega\Delta x \geq 1 $.
%we identify the bound on the maximum momentum ($ \Delta p $) in GUP  with a bound on maximum energy of the system ($ \omega $), i.e. $ \Delta p\geq{1}/{\Delta x}=\omega $. 
Thus, equation (\ref{gup3}) can be written as}
\begin{eqnarray}
\omega_{G}  \geq \omega\left[1 - \frac{\alpha {l_p} }{2(\Delta x)} + \frac{\alpha^{2}{l_p^2}}{2(\Delta x)^{2}}+\cdots\right],
\label{gup5}
\end{eqnarray}
where $\omega$ is the energy of a quantum particle.

Therefore,  for a particle with energy corrected $\omega_{G}$, the tunneling probability reads
\begin{eqnarray}
\Gamma_{G} \cong \exp[-2Im\rm{I_{G}}]. 
\label{gup5.1}
\end{eqnarray} 
Consequently, the corrected temperature becomes
\begin{eqnarray}
T_{HLG} = \frac{\omega}{2 Im \rm{I_{G}}}
\label{gup6}
%\end{eqnarray}
%or better
%\begin{eqnarray}
%T_{HLG} 
= T_{HL}\left[1 - \frac{\alpha{l_p} }{2(\Delta x)} + \frac{\alpha^{2}{l_p^2}}{2(\Delta x)^{2}}+\cdots\right]^{-1}.
\label{gup6.1}
\end{eqnarray}
In following we shall consider three cases.

\subsection{The first case with GUP}
Here we choose $\Delta x = 2x^{+}_{h}= 2\Lambda\sqrt{{2}/{\eta}}$. Thus, for the first case, the corrected temperature due to the GUP is 
\begin{eqnarray}
T_{HLG1} 
%&=& T_{HL1}\left[1 - \frac{\alpha}{4x^{+}_{h}} + \frac{\alpha^{2}}{8x^{+2}_{h}}+\cdots\right]^{-1}\nonumber\\
&=& T_{HL1}\left[1 - \frac{\alpha {l_p} }{4\Lambda}\sqrt{\frac{\eta}{2}} + \frac{\eta\alpha^{2}{l_p^2}}{16\Lambda^2}+\cdots\right]^{-1}.
\label{gup6.2}
\end{eqnarray}
Hence the corrected entropy becomes 
\begin{eqnarray}
S_{HLG1} = \int{\frac{dM}{T_{HLG1}}}
= \left[\pi - \frac{\pi\alpha {l_p}}{4\Lambda}\sqrt{\frac{\eta}{2}}+ \frac{\pi\eta \alpha^{2}{l_p^2}}{16\Lambda^2}
 + \cdots\right]  \ln\left(\frac{M}{M_{0}}\right).
\label{gup8}
\end{eqnarray}
Corrections due to GUP for entropy do not change the dependence of the mass parameter that is always of the type $ \ln(M)$.\\
{\indent As in the previous section, other thermodynamic quantities can also be found. Here they appear to be corrected by the GUP. 
%In the following we shall focus on other important quantities. 
Namely, the corrected specific heat is 
%\begin{eqnarray}
%C_{HLG1} = \pi - \frac{\pi\alpha {l_p}}{4\Lambda}\sqrt{\frac{\eta}{2}}+ \frac{\pi\eta \alpha^{2}{l_p^2}}{16\Lambda^2}
% + \cdots.
%\label{shgup1}
%\end{eqnarray}}
\begin{eqnarray}
C_{HLG1} = C_{HL1}\left[1 - \frac{\alpha {l_p}}{4\Lambda}\sqrt{\frac{\eta}{2}}+ \frac{\eta \alpha^{2}{l_p^2}}{16\Lambda^2}
 + \cdots.\right]
\label{shgup1}
\end{eqnarray}
and the corrected emission rate reads
\begin{eqnarray}
\frac{dM_{HLG1}}{dt} \propto \frac{M^{2}}{\pi^{2}}\left[1 - \frac{\alpha {l_p} }{4\Lambda}\sqrt{\frac{\eta}{2}} + \frac{\eta\alpha^{2}{l_p^2}}{16\Lambda^2}+\cdots\right]^{-2}.
\label{er5}
\end{eqnarray}
Due to the GUP we can also address the issue o minimum mass of black holes. Thus, from the Eq. (\ref{gup2}) we can ensure the following inequality \footnote{We have reinserted a factor such as $\Delta x\to 2\Delta x$ in the GUP.}
\begin{eqnarray}
\label{ineq}
4\alpha^{2}l^{2}_{p} \leq (2\Delta x + \alpha l_{p})^{2}.
\label{mass1}
\end{eqnarray}
%This inequality tells us that the black hole should keep a minimum mass. 
However, in the present case, since the horizon is mass independent, i.e., $\Delta x = 2\Lambda\sqrt{{2}/{\eta}}$, we simply find a minimum length scale given by
\begin{eqnarray}
\Lambda_{min} = \frac{\alpha\, l_{p}}{4\sqrt{2/\eta}}.
\label{length}
\end{eqnarray}
}
%{\color{red} The specific heat for the minimum length scale remain constant.}
\subsection{The second case with GUP}
Now, for the second case we have $\Delta x = 2\Lambda\sqrt{M\Lambda}$ and the corrected temperature due to the GUP reads
\begin{eqnarray}
T_{HLG2} = T_{HL2}\left[1 - \frac{\alpha {l_p}}{4\Lambda\sqrt{M\Lambda}}+ \frac{\alpha^{2}{l_p^2}}{8M\Lambda^3}+\cdots\right]^{-1}.
\label{gup2.1}
\end{eqnarray}
Consequently, for the corrected entropy we obtain
\begin{eqnarray}
S_{HLG2} &=& \int{\frac{dM}{T_{HLG2}}}
%\nonumber\\
= 2\pi\eta\sqrt{M\Lambda} - \frac{\pi\eta\alpha {l_p}}{4\Lambda}\ln\left(\frac{M}{M_0}\right) 
- \frac{\pi\eta\alpha^2 {l_p^2}}{4\Lambda^3}
\sqrt{\frac{\Lambda}{M}}
 + \cdots .
\label{gup12}
\end{eqnarray}
%Notice that we have obtained the quantum corrections to the entropy due to  the effects of GUP considering the tunneling formalism via Hamilton-Jacob method. 
In this example, besides other types of corrections, a logarithmic correction to the entropy of the black hole has been obtained.
{This logarithmic correction arises from the contribution $\alpha l_p(\Delta p)$ in the GUP. }
\\{\indent Again computing other thermodynamic quantities, we have now the corrected specific heat given by
\begin{eqnarray}
C_{HLG2} = 2\pi\eta\sqrt{M\Lambda} - \frac{\pi\eta\alpha {l_p}}{2\Lambda}
+ \frac{\pi\eta\alpha^2 {l_p^2}}{4\sqrt{\frac{\Lambda}{M}}M\Lambda^{2}} + \cdots
\label{shgup12}
\end{eqnarray}
and the corrected emission rate now reads
\begin{eqnarray}
\frac{dM_{HLG2}}{dt} \propto \frac{1}{(\pi\eta)^{2}}\frac{M}{\Lambda}\left[1 - \frac{\alpha {l_p}}{4\Lambda\sqrt{M\Lambda}}+ \frac{\alpha^{2}{l_p^2}}{8M\Lambda^3}+\cdots\right]^{-2}.
\label{er6}
\end{eqnarray}
%Note that the horizon is mass dependent, the Eq. (\ref{gup2}) shown that
%\begin{eqnarray}
%\label{ineq}
%4\alpha^{2}l^{2}_{p} \leq (\Delta x + \alpha l_{p})^{2}.
%\label{mass1}
%\end{eqnarray}
%This inequality tells us that the black hole should keep a minimum mass. 
}
{In the present case the horizon is mass dependent. Thus, substituting $ \Delta x = 2\Lambda\sqrt{M\Lambda} $ into Eq.~(\ref{ineq}), the minimum mass is given by}
\begin{eqnarray}
M_{min} = \frac{\alpha^{2}l_{p}^2}{16\Lambda^{3}}.
\label{mass2}
\end{eqnarray}
%{\bf Thus, the studies of the three cases above suggests that the minimum masses implies the existence of remnants at which the specific heat vanishes.}\\
%{\color{red} At minimum mass limit the specific heat is null.}
\subsection{The AdS-Schwarzschild-like case with GUP}
Now, in this case we have $\Delta x = 2(4M\Lambda^ 4)^{1/3}$ and the corrected temperature due to the GUP reads
\begin{eqnarray}
T_{AdS-S-G} = T_{AdS-S}\left[1 - \frac{\alpha\, {l_p} 4^{2/3}}{16(M\Lambda^4)^{1/3}} + \frac{\alpha^2 {l_p}^2 4^{1/3}}{32(M\Lambda^4)^{2/3}}+\cdots\right]^{-1}.
\label{ads-g1}
\end{eqnarray}
Consequently, for the corrected entropy we obtain
\begin{eqnarray}
S_{AdS-S-G} &=& \int{\frac{dM}{T_{AdS-S-G}}}
%\nonumber\\
%= \frac{4^{2/3}\pi\eta}{3}\ln\left(\frac{M}{M_{0}}\right) - \frac{\pi\eta\alpha {l_p}}{3(4^{2/3})} M 
%+ \frac{\pi\eta\alpha^{2}{l_p^2} }{48} M^{2} + \cdots .
\label{ads-g2}
= {2}^{1/3}\pi \,\eta\,{\left(M\Lambda\right)}^{2/3}-\frac14\,{\frac {\pi \,
\eta\,{2}^{2/3}\alpha\,{\it l_p}\,{\left(M\Lambda\right)}^{1/3}}{{\Lambda }}}+\frac{1}{24}
\,{\frac {\pi \,\eta\,{\alpha}^{2}{{\it l_p}}^{2}
}{{\Lambda}^{2}}}\ln  \left(\frac{M}{M_0}\right) + \cdots 
\end{eqnarray}
%Now, the result has more resemblance with the second case, where contributions other than logarithmic arise. 
{Note that the result has other terms besides the logarithmic contribution. The  $M^{1/3}$ and logarithmic terms come from the $ \alpha l_p (\Delta p) $ and $\alpha^2 l^2_p (\Delta p)^2$ corrections into the GUP, respectively.
%The second term in the above equation is proportional to the mass parameter $ M $ and arises due to the contribution 
%$ \alpha l_p (\Delta p) $ in GUP.
%The last term in the above equation is proportional to the squared mass parameter, which is a similar result to that
%obtained for the black hole entropy in four dimensions. This term is obtained due to the contribution 
%$ \alpha^2 l^2_p (\Delta p)^2 $ in GUP.
}\\
{\indent Finally, in this third case, the corrected specific heat is
\begin{eqnarray}
C_{AdS-S-G} = 2^{\frac{4}{3}}\pi\eta(\Lambda M)^{\frac{1}{2}} - \frac{2^{\frac{2}{3}}}{4}\frac{\pi\eta\alpha l_{p} M}{(M\Lambda)^{\frac{2}{3}}} + \frac{1}{8}\frac{\pi\eta\alpha^{2}l^{2}_{p}}{\Lambda^{2}} + \cdots. 
\label{shads-g2}
\end{eqnarray}
and the corrected emission rate is given by }
\begin{eqnarray}
\frac{dM_{AdS-S-G}}{dt} \propto \frac{9}{4^{4/3}\pi^{2}\eta^{2}}\left(\frac{M}{\Lambda^2}\right)^{2/3}\left[1 - \frac{\alpha\, {l_p} 4^{2/3}}{16(M\Lambda^4)^{1/3}} + \frac{\alpha^2 {l_p}^2 4^{1/3}}{32(M\Lambda^4)^{2/3}}+\cdots\right]^{-2}.\label{er7}
\end{eqnarray}
{In this case replacing  $\Delta x = 2(4M\Lambda^ 4)^{1/3}$ into Eq. (\ref{ineq}) we get the minimum mass }
\begin{eqnarray}
M_{min} = \frac{\alpha^{3}l^{3}_{p}}{256\Lambda^{4}}.
\label{er7-2}
\end{eqnarray}
{As we can easily check, in the three examples studied above, the temperature and emission rate go to zero as $M\to0$, though neither the entropy (unless by considering back reaction effects~\cite{Singleton:2010gz,Singleton:2013ama} in order to address the issue of information loss) nor specific heat vanishes at this limit. However, one has already been shown in the literature that at the minimal mass  the specific heat indeed goes to zero. This is particularly clear as one considers the exact formula of the specific heat \cite{tawfik,Adler:2001vs} rather than the approximated formulas above. By properly working with the GUP we can find an exact expression for the temperature
\begin{equation}\label{temp-cr}
T_{HLG}=2T_H\left(1+\frac{\alpha l_p}{2\Delta x}\right)^{-1}\left[1+\sqrt{1-\frac{4}{\left(1+\frac{2\Delta x}{\alpha l_p}\right)^2}}\right]^{-1}, %\qquad T_H=1/4\pi\Delta x
\end{equation}
which approaches the maximum $T_{max}=T_H$ when saturates the bound (\ref{ineq}), i.e., at $\Delta x=\alpha l_p/2$, where the black hole achieves the minimum mass, as we have discussed above.

As a consequence we can also find an exact expression for the specific heat as follows. Let us first consider the last two cases above, where the horizons depend on black hole mass $M$. Since $T_H=1/4\pi\Delta x$ then making the scaling $\Delta x=\alpha l_p f(M)/2$ into (\ref{temp-cr}), where $f(M)$ is a function of the mass whose first derivative $f'(M)\neq0$, we find the specific heat
\begin{equation}
C_{HLG}=-\Big(1+f(M)\Big)^2\sqrt{\frac{\Big(f(M)+3\Big)\Big(f(M)-1\Big)}{\Big(1+f(M)\Big)^2}}\left(1+ \sqrt{\frac{\Big(f(M)+3\Big)\Big(f(M)-1\Big)}{\Big(1+f(M)\Big)^2}}\right)\frac{\alpha l_p\pi}{f'(M)}.
\end{equation}
Notice that for minimum mass, i.e., as $\Delta x\to\alpha l_p/2$, we have $f(M)\to1$ and then the specific heat $C_{HLG}\to0$.
On the other hand, in the first case above, the entropy (\ref{gup8}) vanishes as $M\to M_0$, where $M_0$ (a minimum mass) is normally associated with the Planck scale. Interesting, the minimum length scale found in (\ref{length}) appears directly related to the Planck length $l_p$, which suggests that $M_0\sim 1/\Lambda_{min}$ is a natural choice.

Thus, the studies of the three cases above show that the minimum masses (or length scale, in the first case) imply the existence of black hole remnants at which the specific heat (or entropy) vanishes, and ceases to radiate even if the effective temperature ($T_{HLG}$) reaches a maximum~\cite{Adler:2001vs}.  In order words, in such a scenario one prevents black holes from entire evaporation \cite{tawfik}. 
%Notice also that the leading contribution of the GUP-corrected thermodynamic quantities agrees with those computed in Sec.~\ref{tunel}.
}\
\section{Product of event horizons}
In this section we will consider the products of horizon.  
{Such products are often formulated in terms of the areas of inner (Cauchy) horizons and outer (event) horizons, and sometimes include the effects of unphysical virtual horizons. }
{It is conjectured that the product of the areas for multi-horizon
stationary black holes are in some cases independent of the mass of the black hole~\cite{Ansorg}.
However, there are studies in the literature where the areas product is dependent on the mass~\cite{Visser}. 
It was also shown in Ref. \cite{Anacleto:2013esa} for acoustic black hole that the universal aspects
of the areas product depends only on quantized quantities such as conserved electric charge and angular momentum.}
Recently, in~\cite{Anacleto:2015kca} has been shown to noncommutative BTZ black holes that the product of
entropy is dependent on the mass parameter $M$ up to linear order in  the non-commutative parameter $\theta$ and becomes independent of the mass when $\theta=0$.
{The areas product with the intriguingly property of depending only on conserved charges has been attracted much interest in string theory \cite{Ansorg} microscopic description of black hole entropy once the area products in terms of quantized charges and quantized angular momenta may provide the basis of  microstates counting.
 In the following we are going to investigate such universal aspects with the introduction of the GUP.}

{Let us first start with the metric (\ref{1.2}) assuming 
$C_{2} = 1/2$, $C_{1} = -M$ and $B = -4M\Lambda^2$, so }
\begin{eqnarray}
x^{\pm}_{h} = \frac{1}{4M} \pm \sqrt{\frac{1}{16M^2} + \frac{2\Lambda^2}{\eta}}.
\end{eqnarray}
Note that
\begin{eqnarray}
x^{+}_{h}x^{-}_{h} =-\frac{2\Lambda^2}{\eta}.
\end{eqnarray}
The product of the radii of horizons is independent of the mass parameter $M$. 

{On the other hand, considering the quantum corrections due to the GUP the horizon radius is changed. From the Eq. (\ref{gup1}) (with $\hbar=1$) and solving for $ \Delta x $ we have
\begin{eqnarray}
\Delta x\geqslant \frac{1}{\Delta p}\left(1 - \alpha{l_{p}}\Delta p + {\alpha^{2}l^{2}_{p}}(\Delta p)^{2}\right),
\end{eqnarray}
which can be written as
\begin{eqnarray}
r_G\geqslant r_h\left(1 - \alpha{l_{p}}M + {\alpha^{2}l^{2}_{p}}M^{2}\right),
\end{eqnarray}
where we have identified $ \Delta x=2r_G $, $ {1}/{\Delta p}=2r_h $ and $ \Delta p= M $. Thus, 
\begin{eqnarray}
x^{\pm}_G\geqslant x^{\pm}_h\left(1 - \alpha{l_{p}}M + {\alpha^{2}l^{2}_{p}}M^{2}\right),
\end{eqnarray}
and the product $ x^{+}_{G}x^{-}_{G} $  becomes 
\begin{eqnarray}
x^{+}_{G}x^{-}_{G} =-\frac{2\Lambda^2}{\eta}\left(1 - \alpha{l_{p}}M + {\alpha^{2}l^{2}_{p}}M^{2}\right)^2.
\end{eqnarray}
This product is now dependent on the mass parameter $ M $.
}

Now we consider the case where $C_{1} = 0$, $C_{2} ={1}/{2}$, $C = 3Q^2\Lambda^2$ and $A = -{1}/{\Lambda^2}$ 
in Eq. (\ref{1.1}) 
\begin{eqnarray}
f(x) = 1 - \frac{1}{\Lambda^2\eta}x^{2} + \frac{B}{\eta x} + \frac{Q^2\Lambda^2}{\eta x^{2}}.
\label{aproximate1}
\end{eqnarray}
At $f(x)=0$ we find the quartic writen as
\begin{eqnarray}
x^{4} - \Lambda^2\left(\eta x^{2} + Bx + Q^2\Lambda^2\right) = 0,
\label{aproximate2}
\end{eqnarray}
or better
\begin{eqnarray}
x^{4} - \Lambda^2\left(x - x_{+}\right)\left(x - x_{-}\right) = 0,
\label{aproximate3}
\end{eqnarray}
where
\begin{eqnarray}
x_{\pm} = \frac{- B \pm \sqrt{B^{2} - 4\eta Q^2\Lambda^2}}{2\eta}.
\label{aproximate9}
\end{eqnarray}

\subsection{Approximate results}
First, we rearrange the quartic to yield the exact equation
\begin{eqnarray}
x = x_{\pm} + \frac{x^{4}}{\Lambda^2(x - x_{\mp})},
\label{aproximate4}
\end{eqnarray}
and then try to solve it perturbatively,
%\subsubsection{Event and Cauchy horizons}
{so for the event horizon we can write the following approximation   }
%To a first approximation, for the event horizon we have
\begin{eqnarray}
x_{E} \approx x_{+} + \frac{x^{4}_{+}}{\Lambda^2(x_{+} - x_{-})} 
= x_{+}\left(1 + \frac{x^{3}_{+}}{\Lambda^2(x_{+} - x_{-})}\right).
\label{aproximate5}
\end{eqnarray}
On the other hand, for the inner (Cauchy) horizon we find
\begin{eqnarray}
x_{C} \approx x_{-} - \frac{x^{4}_{-}}{\Lambda^2(x_{+} - x_{-})} 
= x_{-}\left(1 - \frac{x^{3}_{-}}{\Lambda^2(x_{+} - x_{-})}\right).
\label{aproximate6}
\end{eqnarray}
Consequently the product of horizons is
\begin{eqnarray}
x_{E}x_{C} \approx x_{+}x_{-}\left(1 + \frac{x^{3}_{+} - x^{3}_{-}}{\Lambda^2(x_{+} - x_{-})}\right),
\label{aproximate7}
\end{eqnarray}
or simply 
\begin{eqnarray}
x_{E}x_{C} \approx x_{+}x_{-}\left(1 + \frac{x_{+}^{2}+ x_{+}x_{-} + x_{-}^{2}}{\Lambda^2}\right).
\label{aproximate8}
\end{eqnarray}
In terms of $B$ and $\eta$ we know that
%\begin{eqnarray}
%x_{\pm} = \frac{- B \pm \sqrt{B^{2} + 4\eta}}{2\eta}
%\label{aproximate9}
%\end{eqnarray}
\begin{eqnarray}
x_{+}x_{-} = \frac{Q^2\Lambda^2}{\eta},
\label{aproximate10}
\end{eqnarray}
and
\begin{eqnarray}
x_{\pm}^{2} = \frac{2B^{2} \mp 2B\sqrt{(B^{2}-4\eta Q^2\Lambda^2)} + 4\eta Q^2\Lambda^2}{4\eta^{2}},
\label{aproximate11}
\end{eqnarray}
so that
\begin{eqnarray}
x_{+}^{2}+ x_{+}x_{-} + x_{-}^{2} = \frac{B^{2} + \eta Q^2\Lambda^2}{\eta^{2}}.
\label{aproximate12}
\end{eqnarray}
This implies a product of horizons as follows
\begin{eqnarray}
x_{E}x_{C} \approx \frac{Q^2\Lambda^2}{\eta}\left(1 + \frac{B^{2} - 3\eta Q^2\Lambda^2}{\Lambda^2\eta^{2}}\right).
\label{aproximate13}
\end{eqnarray}
{In addition we also have the relationship}
%For completeness we also note that
\begin{eqnarray}
x_{E} + x_{C} \approx -\frac{B}{\eta} + \frac{x^{4}_{+} - x^{4}_{-}}{\Lambda^2(x_{+} - x_{-})},
\label{aproximate14}
\end{eqnarray}
which again is explicitly dependent on parameters $Q$ and $B$. 
For $\Lambda^2\rightarrow\infty$, identifying $Q$ as the charge and $ B=-4M\Lambda^2 $ the product of the radii of horizons is independent of the mass parameter $M$, i.e. $ x_{E}x_{C} \approx {Q^2\Lambda^2}/{\eta} $,  whereas the sum of the radii is dependent on the mass $M$, $x_{E} + x_{C} \approx {4M\Lambda^2}/{\eta} $.
{In~\cite{Visser} it has been argued that the non-dependence of the mass often fails when the cosmological constant is added to calculate the product of the radii of the horizons.}\\
{\indent Now we consider the quantum corrections due to the GUP for this product. The correction to the equations (\ref{aproximate5}) and (\ref{aproximate6}) are
\begin{eqnarray}
x_{EG}= x_E\left(1 - \alpha{l_{p}}M + {\alpha^{2}l^{2}_{p}}M^{2}\right),
\end{eqnarray}
and 
\begin{eqnarray}
x_{CG}= x_C\left(1 - \alpha{l_{p}}M + {\alpha^{2}l^{2}_{p}}M^{2}\right),
\end{eqnarray}
so for the product $ x_{EG}x_{CG} $ we have obtained
\begin{eqnarray}
x_{EG}x_{CG} = x_{E}x_{C}\left(1 - \alpha{l_{p}}M + {\alpha^{2}l^{2}_{p}}M^{2}\right)^2 \approx \frac{Q^2\Lambda^2}{\eta}\left(1 + \frac{B^{2} - 3\eta Q^2\Lambda^2}{\Lambda^2\eta^{2}}\right)\left(1 - \alpha{l_{p}}M + {\alpha^{2}l^{2}_{p}}M^{2}\right)^2.
\label{aproximategup1}
\end{eqnarray}
}
{For the sum of the radii we have}
\begin{eqnarray}
x_{EG} + x_{CG} = \left(x_{E} + x_{C}\right)\left[1 - \alpha{l_{p}}M + {\alpha^{2}l^{2}_{p}}M^{2}\right] \approx \left(-\frac{B}{\eta} + \frac{x^{4}_{+} - x^{4}_{-}}{\Lambda^2(x_{+} - x_{-})}\right)\left[1 - \alpha{l_{p}}M + {\alpha^{2}l^{2}_{p}}M^{2}\right].
\label{aproximategup2}
\end{eqnarray}

 Before finishing this section let us address the issues concerning cosmological and and virtual horizons as in the following.
\subsubsection{Cosmological horizon}

Let us go back to the quartic to write it as in the following 
\begin{eqnarray}
x^{2} = \frac{\Lambda^2(x-x_{+})(x-x_{-})}{x^{2}}.
\label{aproximate15}
\end{eqnarray}
{For a zero-order approximation we have}
%where we have a zero order
\begin{eqnarray}
x_{CH} = \Lambda,
\label{aproximate16}
\end{eqnarray}
{and that for a first order of approximation gives}
\begin{eqnarray}
x_{CH} &=& \Lambda\sqrt{\frac{(\Lambda - x_{+})(\Lambda - x_{-})}{\Lambda^2}} 
%\nonumber\\
%&=& 
=\Lambda\left(1 - \frac{x_{+} + x_{-}}{2\Lambda}+\cdots \right),
\nonumber\\
&=& \Lambda\left(1 + \frac{B}{2\eta\Lambda}+\cdots\right).
\label{aproximate17}
\end{eqnarray}
{Hence the cosmological horizon becomes}
\begin{eqnarray}
x_{CH} = \Lambda + \frac{B}{2\eta} .
\label{aproximate18}
\end{eqnarray}
{Thus, in this order of approximation, the result for the cosmological horizon does not depend on $ Q $, but depends on $ B = -4M\Lambda^2$, and consequently depends on the mass parameter $ M $.}\\
{The corrected cosmological horizon is
\begin{eqnarray}
x_{CHG} = x_{CH}\left[1 - \alpha{l_{p}}M + {\alpha^{2}l^{2}_{p}}M^{2}\right] =  \left(\Lambda + \frac{B}{2\eta}\right)\left[1 - \alpha{l_{p}}M + {\alpha^{2}l^{2}_{p}}M^{2}\right] .
\label{aproximategup3}
\end{eqnarray}

}
\subsubsection{Virtual horizon}
{Therefore, considering the exact result we have the following virtual horizon}
%Finally, from the exact quartic we find
\begin{eqnarray}
x_{V} = - (x_{E} + x_{C} + x_{CH})=-\Lambda- \frac{B}{\eta},
\label{aproximate19}
\end{eqnarray}
%so that
%\begin{eqnarray}
%x_{V} = -\sqrt{a} - \frac{B}{\eta}.
%\label{aproximate20}
%\end{eqnarray}
{that is also dependent on the mass parameter $ M $}.\\
{
\indent The corrected virtual horizon is
\begin{eqnarray}
x_{VG} = x_{V}\left[1 - \alpha{l_{p}}M + {\alpha^{2}l^{2}_{p}}M^{2}\right] =  \left(-\Lambda - \frac{B}{\eta}\right)\left[1 - \alpha{l_{p}}M + {\alpha^{2}l^{2}_{p}}M^{2}\right] .
\label{aproximategup4}
\end{eqnarray}}

{Notice that the effect of the GUP is essentially to reveal a mass dependence of the event horizon products. As we previously mentioned, a similar role is played by non-commutative black holes. This effect at least shows more examples where the the conjectured mass independence of the event horizon products
often fails once either a cosmological constant is added or quantum corrections via GUP are present. 
%However, things can change as long as the mass $M\to0$, as an evaporation effect.
%leaving behind a minimal product left that can be associated with the entropy of extremal black holes.
}

\section{Conclusions}
In summary, by considering the GUP, we derive the two-dimensional Horava-Lifshitz black hole temperature and entropy using the Hamilton-Jacobi version of the tunneling formalism. In our calculations the Hamilton-Jacobi method was applied to calculate the imaginary part of the action and the GUP was introduced by the correction to the energy of a particle due to gravity near horizon.  We apply this to essentially three types of black holes: Schwarzschild, Reissner-Nordstr\"om and AdS-Schwarzschild-like black holes. Furthermore, we also addressed the issues of event horizon products that can find relevance in computation of the microscopic entropy in a $AdS_2/CFT_1$ correspondence. Several other interesting issues, such as applying holography techniques in two-dimensional models of QCD by using 2d AdS-Schwarzschild-like black holes should be addressed elsewhere. 
%The unusual positivity of the coefficients of the logarithm terms in equation (\ref{gup12}) gives a stronger contribution to the entropy.

\section*{Acknowledgement}
We would like to thank CAPES and CNPq for financial support.


\begin{thebibliography}{00}
\bibitem{mann} R. B. Mann, A. Shiekh and L. Tarasov, Nucl. Phys. B341, 134 (1990).

\bibitem{twodimentional} C.G. Callan, S.B. Giddings, J.A. Harvey, and A. Strominger, Phys. Rev. D45 (1992) 1005(arXiv:hepth/
9111056v1); A. E. Sikkema and R. B Mann, Class. Quantum Grav. 8 (1991) 219; D. Christensen and R. B. Mann, Class. Quant. Grav. 9, 1769 (1992) [hep-th/9203050]; C. Eling and T. Jacobson, Phys. Rev. D 74, 084027 (2006) [gr-qc/0608052]; S. M. Christensen and S. A. Fulling, Phys. Rev. D 15, 2088 (1977); S. P. Trivedi, Phys. Rev. D 47, 4233 (1993).

\bibitem {Polyakov87}  A. M. Polyakov, Mod. Phys. Lett. A2,893 (1987).

\bibitem{Teitelboim} C. Teitelboim, in Quantum theory of gravity, ed. S. Christensen (Adam Hilger, Bristol, 1984) p.
327;
R. Jackiw, in Quantum theory of gravity, ed. S. Christensen (Adam Hilger, Bristol, 1984) p. 403;
NucI. Phys. B252 (1985) 343.

\bibitem{Polyakov88} V. Kniznik, AM. Polyakov and A.B. Zamolodchikov, Mod. Phys. Lett. A3, 819 (1988) .

\bibitem{leblanc88} M. Leblanc, R.B. Mann and B. Shadwick, Phys. Rev. D {\bf 37} (1988) 3548
J. Gegenberg, P.F. Kelly, RB. Mann, R. McArthur and D.E. Vincent, Mod. Phys. Lett. A3 (1988) 1791;
 J. Gegenberg, P.F. Kelly, G. Kunstatter, RB. Mann, R. McArthur and D.E. Vincent, Phys. Rev. D {\bf 40}, 1919 (1989);
U. Lindström and M. Roèek, Class. Quant. Gray. 4 (1987) L79.

\bibitem{Horava} P. Horava, Phys. Rev. D {\bf 79}, 084008 (2009) [arXiv:0901.3775 [hep-th]].

\bibitem{horava2} M. Visser, Phys. Rev. D {\bf 80}, 025011 (2009); T. Sotiriou, M. Visser, and S. Weinfurtner, J. High Energy Phys. 10, 033 (2009); C. Bogdanos, E.N. Saridakis, Class. Quant. Grav. 27, 075005 (2010); A. Wang and R. Maartens, Phys. Rev. D {\bf 81}, 024009 (2010); Y.-Q. Huang, A. Wang, and Q. Wu, Mod. Phys. Lett. A25, 2267 (2010); C. Charmousis, G. Niz, A. Padilla, and P.M. Saffin, J. High Energy Phys. 08, 070 (2009); D. Blas, O. Pujolas, and S. Sibiryakov, J. High Energy Phys. 10, 029 (2009); K. Koyama and F. Arroja, J. High Energy Phys. 03, 061 (2010); I. Kimpton and A. Padilla, J. High Energy Phys. 07, 014 (2010); A. Wang and Q. Wu, Phys. Rev. D {\bf 83}, 044025 (2011); B. Chen and Q. G. Huang, Phys.
Lett. B 683, 108 (2010) [arXiv:0904.4565 [hep-th]].

\bibitem{geraldo} D. Bazeia, F. A. Brito and F. G. Costa, Phys.Rev.D.91, (2015) 044026.
 

%\bibitem{Bekenstein} J. Bekenstein, Phys. Rev. D7 (1973) 2333; Phys. Rev. D9 (1974) 3292; Lett. Nuovo Cimento 4 (1972) %737.

%%%%%%%%%%%%%%%%%%%%%%%%%%%%%%%%%%%%%%%%%%%%%%%%%%%%%%%%%%%%%%%

\bibitem{Parikh} M. K. Parikh and F. Wilczek, Phys. Rev. Lett. {\bf 85}, 5042 (2000).
%\bibitem{mk.parikh-prl85} 
 % Parikh, M.K. Wilczek, F.,
  %Phys. Rev. Lett. {\bf 85}, 5042 (2000). 
%\bibitem{mk.parikh-hepth0402166} 
 M. K. Parikh, hep-th/0402166. 

\bibitem{ec.vagenas-plb559} Vagenas, E.C., Phys. Lett. B {\bf 559}, 65 (2003).


\bibitem{SP} K. Srinivasan and T. Padmanabhan, Phys. Rev. D {\bf 60}, 024007 (1999), [arXiv:gr-qc/9812028]; 
S. Shankaranarayanan, K. Srinivasan and T. Padmanabhan, Mod. Phys. Lett. A {\bf 16}, 571 (2001), [arXiv:gr-qc/0007022]; S. Shankaranarayanan, T. Padmanabhan and K. Srinivasan, Class. Quantum Grav. {\bf 19}, 2671 (2002), [arXiv:gr-qc/0010042]. E. T. Akhmedov et al. , Int. J. Mod. Phys. A{\bf22} (2007) 1705-1715; [hep-th/0605137]
E. T. Akhmedov et al., Phys. Lett. B{\bf642} (2006) 124-128; [hep-th/0608098]. V. Akhmedova et al.; Phys. Lett. B{\bf666} (2008) 269-271; [arXiv:0804.2289 [hep-th]]. E.T. Akhmedov et al.; Int. J. Mod. Phys. D{\bf17} (2008) 2453-2458; [arXiv:0805.2653 [gr-qc]]. V. Akhmedova et al.;  Phys. Lett. B{\bf673} (2009) 227-231; [arXiv:0808.3413 [hep-th]]. 

%\cite{Eune:2010kx}
\bibitem{Eune:2010kx} 
  M.~Eune and W.~Kim,
  %``Lifshitz scalar, brick wall method, and GUP in Horava-Lifshitz Gravity,''
  Phys.\ Rev.\ D {\bf 82}, 124048 (2010)
  doi:10.1103/PhysRevD.82.124048
  [arXiv:1007.1824 [hep-th]];
  %%CITATION = doi:10.1103/PhysRevD.82.124048;%%
  %8 citations counted in INSPIRE as of 10 Mar 2016
  
De-You Chen,  Haitang Yang, Xiao-Tao Zu, Phys. Lett. B {\bf681} (2009) 463"1¤7468; B.~R.~Majhi, Phys. Lett. B{\bf686}:49 (2010).
  
  %\cite{Liu:2011gya}
\bibitem{Liu:2011gya} 
  M.~Liu, J.~Lu and J.~Lu,
  %``Fermions Analysis of IR modified Horava-Lifshitz gravity: Tunneling and Perturbation Perspectives,''
  Class.\ Quant.\ Grav.\  {\bf 28}, 125024 (2011)
  doi:10.1088/0264-9381/28/12/125024
  [arXiv:1108.0758 [hep-th]];
  %%CITATION = doi:10.1088/0264-9381/28/12/125024;%%
  %7 citations counted in INSPIRE as of 10 Mar 2016
Zeng Xiao-Xiong,  Li Ling, Commun. Theor. Phys. {\bf55} (2011) 376"1¤7380.

\bibitem{Jiang} Q. -Q. Jiang, S. -Q. Wu and X. Cai, Phys. Rev. D {\bf 73}, 064003 (2006); Z. Xu and B.
Chen, Phys. Rev. D {\bf 75}, 024041 (2007).

\bibitem{Banerjee} R. Banerjee and B. R. Majhi, Phys. Lett. B {\bf 662}, 62 (2008).

%\cite{Singleton:2010gz}
\bibitem{Singleton:2010gz} 
  D.~Singleton, E.~C.~Vagenas, T.~Zhu and J.~R.~Ren,
  %``Insights and possible resolution to the information loss paradox via the tunneling picture,''
  JHEP {\bf 1008}, 089 (2010)
  Erratum: [JHEP {\bf 1101}, 021 (2011)]
  doi:10.1007/JHEP08(2010)089, 10.1007/JHEP01(2011)021
  [arXiv:1005.3778 [gr-qc]].
  
%\cite{Singleton:2013ama}
\bibitem{Singleton:2013ama} 
  D.~Singleton, E.~C.~Vagenas and T.~Zhu,
  %``Self-similarity, conservation of entropy/bits and the black hole information puzzle,''
  JHEP {\bf 1405}, 074 (2014)
  doi:10.1007/JHEP05(2014)074
  [arXiv:1311.2015 [gr-qc]].  

\bibitem{Silva12} 
  C.~A.~S.~Silva and F.~A.~Brito,
  %``Quantum tunneling radiation from self-dual black holes,''
  Phys.\ Lett.\ B {\bf 725}, no. 45, 456 (2013)
  [arXiv:1210.4472 [physics.gen-ph]].
  %%CITATION = ARXIV:1210.4472;%%
  
   
  
  %\cite{Majumder:afa}
\bibitem{Majumder:afa} 
  B.~Majumder,
  %``Black Hole Entropy with minimal length in Tunneling formalism,''
  Gen.\ Rel.\ Grav.\  {\bf 11}, 2403 (2013)
  [arXiv:1212.6591 [gr-qc]].
  %%CITATION = ARXIV:1212.6591;%%

%\cite{Becar:2010zza}
\bibitem{Becar:2010zza} 
M.~A.~Anacleto, F.~A.~Brito, G.~C.~Luna, E.~Passos and J.~Spinelly,
 Annals Phys.\  {\bf 362}, 436 (2015)
 [arXiv:1502.00179 [hep-th]];
  R.~Becar, P.~Gonzalez, G.~Pulgar and J.~Saavedra,
  %``Hawking radiation via anomaly and tunneling method by Unruh's and canonical acoustic black hole,''
  Int.\ J.\ Mod.\ Phys.\ A {\bf 25}, 1463 (2010).
  %%CITATION = IMPAE,A25,1463;%%
M. A. Anacleto, F. A. Brito, E. Passos, Phys. Rev. D {\bf 87}, 125015 (2013) [ arXiv:1210.7739 ]; Phys. Rev. D {\bf 85}, 025013 (2012) [arXiv:1109.6298 [hep-th]], 
M. A. Anacleto, F. A. Brito, E. Passos, Phys. Lett. B {\bf694}, 149 (2010)  [arXiv:1004.5360 [hep-th]]; Phys. Lett. B {\bf703}, 609 (2011) [arXiv:1101.2891 [hep-th]];
Phys. Rev. D {\bf 86}, 125015 (2012)  [arXiv:1208.2615 [hep-th]].
%{\it Acoustic Black Holes and Universal Aspects of Area Products }  [arXiv:1309.1486] .

%\cite{Anacleto:2015mma}
\bibitem{Anacleto:2015mma} 
  M.~A.~Anacleto, F.~A.~Brito and E.~Passos,
  %``Quantum-corrected self-dual black hole entropy in tunneling formalism with GUP,''
  Phys.\ Lett.\ B {\bf 749}, 181 (2015)
  [arXiv:1504.06295 [hep-th]].

%\cite{Anacleto:2015kca}
\bibitem{Anacleto:2015kca} 
  M.~A.~Anacleto, F.~A.~Brito, A.~G.~Cavalcanti, E.~Passos and J.~Spinelly,
  %``Quantum correction to the entropy of noncommutative BTZ black hole,''
  arXiv:1510.08444 [hep-th].

%\cite{Ovgun:2015box}
\bibitem{Ovgun:2015box} 
  A. \"Ovg\"un and K.~Jusufi,
  %``Massive Vector Particles Tunneling From Noncommutative Charged Black Holes,''
  arXiv:1512.05268 [gr-qc].

\bibitem{Faizal:2014tea} 
  M.~Faizal and M.~M.~Khalil,
  %``GUP-Corrected Thermodynamics for all Black Objects and the Existence of Remnants,''
  Int.\ J.\ Mod.\ Phys.\ A {\bf 30}, no. 22, 1550144 (2015)
  [arXiv:1411.4042 [gr-qc]].

\bibitem{Wilczek} V. Frolov, I. Novikov, Phys. Rev. D {\bf 48}, 4545 (1993) ; 
C.G. Callan, F. Wilczek, Phys. Lett. B {\bf 33}, 55 (1994).


%\cite{Magan:2014dwa}
\bibitem{Magan:2014dwa} 
  J.~M.~Mag\'an, D.~Melnikov and M.~R.~O.~Silva,
  %``Black Holes in AdS/BCFT and Fluid/Gravity Correspondence,''
  JHEP {\bf 1411}, 069 (2014)
  [arXiv:1408.2580 [hep-th]];
  %%CITATION = ARXIV:1408.2580;%%
  H.~Casini, M.~Huerta and R.~C.~Myers,
  %``Towards a derivation of holographic entanglement entropy,''
  JHEP {\bf 1105}, 036 (2011)
  [arXiv:1102.0440 [hep-th]].

  
  %\cite{Solodukhin:2011gn}
\bibitem{Solodukhin:2011gn} 
  S.~N.~Solodukhin,
  %``Entanglement entropy of black holes,''
  Living Rev.\ Rel.\  {\bf 14}, 8 (2011)
  [arXiv:1104.3712 [hep-th]].
  %%CITATION = ARXIV:1104.3712;%%

\bibitem{mhorizon}
Wei Xu, Jia Wang, Xin-he Meng, Galaxies 2015, 3(1), 53-71, 
[arXiv:1402.1293 [hep-th]];
Wei Xu, Jia Wang, Xin-he Meng, 
Int. J. Mod. Phys. A {\bf 29} (2014) 18, 1450088,
[arXiv:1401.5180 [gr-qc]];
Jia Wang, Wei Xu, Xin-he Meng, 
Phys. Rev. D {\bf 89}, 044034 (2014) , 
[arXiv:1312.3057 [gr-qc]];
Wei Xu, Jia Wang, Xin-he Meng, Int.\ J.\ Mod.\ Phys.\ A {\bf 29}, 1450172 (2014),
[arXiv:1310.7690 [gr-qc]];
Jia Wang, Wei Xu, Xin-He Meng
JHEP {\bf 1401}, 031 (2014), 
[arXiv:1310.6811 [gr-qc]].



\bibitem{Kaul} R. K. Kaul and P. Majumdar, Phys. Rev. Lett. {\bf 84},  5255 (2000);
R. K. Kaul and P. Majumdar, Phys. Lett. B {\bf 439}, 267 (1998).     

%\cite{Carlip:2000nv}
\bibitem{Carlip:2000nv} 
  S.~Carlip,
  %``Logarithmic corrections to black hole entropy from the Cardy formula,''
  Class.\ Quant.\ Grav.\  {\bf 17}, 4175 (2000)
  [gr-qc/0005017].
  
\bibitem{Rinaldi}  S. Giovanazzi, Phys. Rev. Lett. 106, 011302 (2011), M. Rinaldi, Phys. Rev. D {\bf 84}, 124009 (2011).
%\bibitem{Rinaldi:2011aa}
   M.~Rinaldi,
   %``The entropy of an acoustic black hole in Bose-Einstein condensates: transverse
%modes as a cure for divergences,''
   Int.\ J.\ Mod.\ Phys.\ D {\bf 22} (2013) 1350016
   [arXiv:1112.3596 [gr-qc]].  
  
\bibitem{Brustein}   Ram Brustein, and Judy Kupferman, 
%"Black hole entropy divergence and the uncertainty principle", 
Phys. Rev. D {\bf 83}, 124014 (2011), [arXiv:1010.4157 [hep-th]]; 
Kim, Wontae, Kim, Yong-Wan, and Park, Young-Jai, 
%?Entropy of 2+1 de Sitter space with the GUP?, 
J. Korean Phys. Soc., 49, 1360, (2006), [arXiv:gr-qc/0604065 [gr-qc]]; 
Kim, Yong-Wan, and Park, Young-Jai, 
%?Entropy of the Schwarzschild black hole to all orders in the Planck length?, 
Phys. Lett., B {\bf 655}, 172 (2007), 
%\href{run:arXiv.org/abs/0707.2128}
[arXiv:0707.2128 [gr-qc]]; 
Sun, Xue-Feng, and Liu, Wen-Biao, 
%?Improved black hole entropy calculation without cutoff?,
Mod. Phys. Lett., A {\bf 19}, 677 (2004); 
Yoon, Myungseok, Ha, Jihye, and Kim, Wontae, 
%?Entropy of Reissner-Nordstrom Black Holes with Minimal Length Revisited?, 
Phys. Rev., D {\bf 76}, 047501 (2007), 
%\href{run:arXiv.org/abs/0706.0364}
[arXiv:0706.0364 [gr-qc]].

\bibitem{XLi} X. Li, Z. Zhao, Phys. Rev. D {\bf 62}, 104001 (2000);
R. Zhao, S. L. Zhang, Gen. Relat. Grav. {\bf 36}, 2123 (2004);
R. Zhao, Y. Q. Wu, L. C. Zhang, Class. Quantum Grav. {\bf 20}, 4885 (2003);
W. Kim, Y. W. Kim, Y. J. Park, Phys. Rev. D {\bf 74}, 104001 (2006); Phys. Rev. D {\bf 75}, 127501 (2007);
M. Yoon, J. Ha, W. Kim, Phys. Rev. D {\bf 76},  047501 (2007);
Y. W. Kim, Y. J. Park, Phys. Lett. B {\bf 655}, 172 (2007).

\bibitem{KN} K. Nouicer, Phys. Lett. B {\bf 646}, 63 (2007).

\bibitem{ABPS} 
  M.~A.~Anacleto, F.~A.~Brito, E.~Passos and W.~P.~Santos,
  %``The entropy of the noncommutative acoustic black hole based on generalized uncertainty principle,''
  Phys.\ Lett.\ B {\bf 737}, 6 (2014)
 % \href{run:arxiv.org/abs/1405.2046}
  [arXiv:1405.2046 [hep-th]].
  %%CITATION = ARXIV:1405.2046;%%
\bibitem{Zhao} HuiHua Zhao, GuangLiang Li, LiChun Zhang, Phys.Lett. A {\bf 376},  2348 (2012).

\bibitem {Zhang} R. Zhao, L.C. Zhang, H.F. Li, Acta Phys. Sin. 58,  2193 (2009) . 

\bibitem{review} A.~Nasser Tawfik and A.~Magied Diab,
  %``Black Hole Corrections due to Minimal Length and Modified Dispersion Relation,''
  Int.\ J.\ Mod.\ Phys.\ A {\bf 30}, no. 12, 1550059 (2015)
  doi:10.1142/S0217751X15500591
  [arXiv:1502.04562 [gr-qc]].
  

%\cite{Adler:2001vs}
\bibitem{Adler:2001vs} 
  R.~J.~Adler, P.~Chen and D.~I.~Santiago,
  %``The Generalized uncertainty principle and black hole remnants,''
  Gen.\ Rel.\ Grav.\  {\bf 33}, 2101 (2001)
  doi:10.1023/A:1015281430411
  [gr-qc/0106080];
  %\cite{Adler:1999bu}
%\bibitem{Adler:1999bu} 
  R.~J.~Adler and D.~I.~Santiago,
  %``On gravity and the uncertainty principle,''
  Mod.\ Phys.\ Lett.\ A {\bf 14}, 1371 (1999)
  doi:10.1142/S0217732399001462
  [gr-qc/9904026].

%\cite{Nicolini:2008aj}
\bibitem{Nicolini:2008aj} 
  P.~Nicolini,
  %``Noncommutative Black Holes, The Final Appeal To Quantum Gravity: A Review,''
  Int.\ J.\ Mod.\ Phys.\ A {\bf 24}, 1229 (2009)
  doi:10.1142/S0217751X09043353
  [arXiv:0807.1939 [hep-th]];
%\cite{Nicolini:2005vd}
%\bibitem{Nicolini:2005vd} 
  P.~Nicolini, A.~Smailagic and E.~Spallucci,
  %``Noncommutative geometry inspired Schwarzschild black hole,''
  Phys.\ Lett.\ B {\bf 632}, 547 (2006)
  doi:10.1016/j.physletb.2005.11.004
  [gr-qc/0510112].  
  

%\cite{Gangopadhyay:2015zma}
\bibitem{Gangopadhyay:2015zma} 
  S.~Gangopadhyay, A.~Dutta and M.~Faizal,
  %``Constraints on the Generalized Uncertainty Principle from Black Hole Thermodynamics,''
  Europhys.\ Lett.\  {\bf 112}, no. 2, 20006 (2015)
  doi:10.1209/0295-5075/112/20006
  [arXiv:1501.01482 [gr-qc]].  

\bibitem{Bekenstein} J. Bekenstein, Phys. Rev. D7 (1973) 2333; Phys. Rev. D9 (1974) 3292; Lett. Nuovo Cimento 4 (1972) 737.

\bibitem{tawfik} R. Emparan, G. T. Horowitz and R. C. Myers, Phys. Rev. Lett. 85, 499 (2000),
[arXiv:0003118[hep-th]]; A. Tawfik, JCAP 07 (2013) 040, [arXiv:1307.1894 [gr-qc]].


\bibitem{Ansorg} M. Ansorg and J. Hennig, Class. Quant. Grav. {\bf 25}, 222001  (2008), 
[arXiv:0810.3998 [gr-qc]];
Phys. Rev. Lett. {\bf 102}, 221102 (2009) [arXiv:0903.5405 [gr-qc]];
J. Hennig and M. Ansorg, Annales Henri Poincare {\bf 10} (2009) 1075 [arXiv:0904.2071 [gr-qc]];
M. Ansorg, J. Hennig, and C. Cederbaum,  Gen. Rel. Grav. {\bf 43} (2011) 1205 [arXiv:1005.3128 [gr-qc]];
M. Cvetic, G. W. Gibbons, and C. N. Pope,  Phys. Rev. Lett. {\bf 106} (2011) 121301 [arXiv:1011.0008 [hep-th]];
A. Castro and M. J. Rodriguez, Phys. Rev. D {\bf 86} (2012) 024008 [arXiv:1204.1284 [hep-th]];
M. Cvetic, H. Lu, and C. N. Pope, Phys.\ Rev.\ D {\bf 88}, 044046 (2013), arXiv:1306.4522 [hep-th].

\bibitem{Visser} M. Visser, Phys. Rev. D {\bf 88}, 044014 (2013), [arXiv:1205.6814 [hep-th]].

\bibitem{Anacleto:2013esa} 
  M.~A.~Anacleto, F.~A.~Brito and E.~Passos,
  %``Acoustic Black Holes and Universal Aspects of Area Products,''
  Phys.\ Lett.\ A {\bf 380}, 1105 (2016)
  doi:10.1016/j.physleta.2016.01.030
  arXiv:1309.1486 [hep-th].


\end{thebibliography}
\end{document}